\documentclass[showpacs, preprintnumbers, amsmath, amssymb,twocolumn]{revtex4}

\usepackage{graphicx}
\usepackage{dcolumn}
\usepackage{bm}
\usepackage{subfigure}

\begin{document}

\preprint{...}

\title{The interference and gravitational redshift effect of long waves passing a binary black hole}

\author{Qiyun Fu, Tieyan Si}
\affiliation{School of Physics, Harbin Institute of Technology, Harbin, 150080, China}

\date{\today}

\begin{abstract}

We investigate the interference of electromagnetic long waves passing a binary black hole based on the approximate binary black hole metric. The interference pattern of long waves demonstrates strong contrast intensity and changes with respect to different wavelengths and incoming angles. A bright semicircular arc emerges from the interference pattern and bridges the two black holes when the binary black hole rotates to certain angle. The angular momentum of the binary black hole causes asymmetric gravitational redshift distribution along the relative position vector of the two black holes. The angular momentum of the binary black hole is measurable based on the interference pattern of long waves and gravitational redshift.

\end{abstract}


\maketitle

\section{Introduction}

A moving particle (including photon) in vicinity of a massive celestial body travels along a curved geodesic path due to the highly curved spacetime. The strong gravitational field acts as a gravitational lens, focusing many photons along curved geodesic paths onto one point. The shape of light source is usually highly distorted by gravitational lens, sometimes projects multiple images of the same light sources \cite{0-6,0-7}. A ring of light sources named after Einstein ring is observed when an observer, a gravitational lens and a light source are aligned in a straight line \cite{0-1,0-2,0-3}. In most cases, the observer, gravitational lens and light source are not in a straight line, the image of light source is divided into multiple copies \cite{0-4,0-4-1}, and highly amplified by the convergence of light \cite{0-5}. When the wavelength of electromagnetic wave is comparable to the Schwarzschild radius of the gravitational lens, the distant light source is not distinguishable by the telescope, the Einstein ring is no longer observable \cite{1-1-1,1-1-2,1-2}. In that case, the interference, diffraction or reflection effect of this long wave becomes the dominant physical effect.

Besides the long-wavelength electromagnetic waves, the gravitational wave produced by merging two black holes (or two massive neutron stars) also has a long wavelength comparable to the distance between the binary stars, leading to apparat gravitational lensing effect \cite{1-3,1-4,1-5,1-6,1-7,1-8}. In the gravitational lensing effect of binary stars, one of the binary stars acts as a radiation source for electromagnetic waves, while the other star acts as a gravitational lens that converges electromagnetic waves \cite{2-1,2-1-2,2-1-3}. The wave optical effects of gravitational lends provides a basic method for the study of exoplanets \cite{2-2}.

The orbital velocity of the binary-black holes around their mass center is large enough to distort the light path away from classical optical path \cite{2-3}. We apply the numerical method of ray tracing which was used for simulating the electromagnetic field distribution around binary black hole\cite{3-2}, to find the convergence path of waves in vicinity of binary black hole. Similar to the gravitational redshift caused by single black hole, the electromagnetic waves emitted by the surrounding object around the binary black hole also has gravitational shift in the perpendicular direction to the rotating plane \cite{3-3,3-3-2,3-3-3,3-4}.

Here we study the gravitational redshift within the rotating plane of the binary black hole. The directed rotation of binary black hole breaks spherical symmetry of single black hole and results in asymmetric trajectories of waves on the left hand and the right hand side of the binary black hole, leading to asymmetric gravitation redshift effect. The propagation paths of lights in binary black hole spacetime are traced numerically by geodesic equation. Parallel waves out of the same source interferant with one another when they intersect at the same point. Unlike the constant frequency of waves in Euclidean space time, the frequency of wave in the space time of the binary black hole is no longer a constant, the gravitational redshift has strong influence on the interference patterns of intersecting waves. The articles is organized into two sections: in section II, the diffraction and interference of long waves in vicinity of the binary black hole is investigated. In section III, we study the asymmetric distribution of gravitational redshift around the binary black hole.

\section{The diffraction and interference of long waves in vicinity of the binary black hole}
\label{sec:interference}

For two Schwarzschild black holes that are far apart but orbiting around each other along circular tracks, we focus on the wave propagation confined in the orbital plane of the binary black holes by an approximated metric solution of binary black hole \cite{Alvi}. The center of mass of binary black hole is located at the origin of the coordinates. The projected metric $g^{\mu\nu}$ of the binary black hole metric to the orbital plane is expressed by Cartesian coordinate $({t,x,y})$,
\begin{eqnarray}
g_{00}&=&-1+\frac{2m_1}{r_1}+\frac{2m_2}{r_2}
-2\left(\frac{m_1}{r_1}+\frac{m_2}{r_2}\right)^2\nonumber\\
&+&\frac{m_1}{r_1}\left[4v_1^2-(\textbf{n}_1\cdot \textbf{v}_1)^2\right]
+\frac{m_2}{r_2}\left[4v_2^2-(\textbf{n}_2\cdot \textbf{v}_2)^2\right]\nonumber\\
&-&2\frac{m_1m_2}{b}\left(\frac{1}{r_1}+\frac{1}{r_2}\right)
+\frac{m_1m_2}{b^3}\textbf{b}\cdot (\textbf{n}_1-\textbf{n}_2),\nonumber\\
g_{0i}&=&-4\left(\frac{m_1}{r_1}v_1^i+\frac{m_2}{r_2}v_2^i\right),\nonumber\\
g_{ij}&=&\delta_{ij}\left(1+\frac{2m_1}{r_1}+\frac{2m_2}{r_2}\right).
\end{eqnarray}
The initial locations of the two block holes at $t=0$ are $\boldsymbol{x}_1=(-b/2,0,0)$ and $\boldsymbol{x}_2 = (b/2,0,0)$. $r_q$ is the distance between a location point and the $q$th black hole,
\begin{eqnarray}
r_q&=&|\boldsymbol{x}-\boldsymbol{x}_q|, \;\;\;\boldsymbol{n}_q=\frac{\boldsymbol{x}-\boldsymbol{x}_q}{r_q},\nonumber\\
\boldsymbol{v}_q &=&\frac{\mathrm{d}\boldsymbol{x}_q}{\mathrm{d}t}, \;\;\; v_q=|\boldsymbol{v}_q|,\;\; q=1,2.
\end{eqnarray}
$\boldsymbol{v}_q $ is the velocity of the $q$th black hole. The geodesic path of photon in vicinity of black hole is derived from the geodesic equation,
\begin{eqnarray}\label{geodesic}
   \frac{\text{d}^2x^{\mu}}{\text{d}s^2} &+&\Gamma _{\nu \sigma}^{\mu}\frac{\text{d}x^{\nu}}{\text{d}s}  \frac{\text{d}x^{\sigma}}{\text{d}s}=0,\nonumber\\
  \Gamma^\alpha_{\lambda\mu} &=&\frac12g^{\alpha\nu}(g_{\mu\nu,\lambda}+g_{\nu\lambda,\mu}-g_{\lambda\mu,\nu}).
\end{eqnarray}
The geodesic equation is reexpressed as a first-order ordinary differential equation by introducing $u^{\mu}=\text{d}x^{\mu}/\text{d}s$,
\begin{eqnarray}\label{geodesicODE}
  \frac{\text{d}x}{\text{d}t}&=&\frac{u_x}{u_t},\;\;\;\frac{\text{d}y}{\text{d}t}=\frac{u_y}{u_t},\;\;\;
  \frac{\text{d}u_t}{\text{d}t}=
  -\frac{1}{u_t}\Gamma _{\nu \sigma}^{0}u^{\nu}u^{\sigma},\nonumber\\
  \frac{\text{d}u_x}{\text{d}t}&=&
  -\frac{1}{u_t}\Gamma _{\nu \sigma}^{1}u^{\nu}u^{\sigma},\;\;\;\;\;
  \frac{\text{d}u_y}{\text{d}t}=
  -\frac{1}{u_t}\Gamma _{\nu \sigma}^{2}u^{\nu}u^{\sigma}.
\end{eqnarray}
The geodesic path is derived from the numerical solution of [$x{(t)}$, $y{(t)}$] in Eq. (\ref{geodesicODE}).

\begin{figure}[ht]
\begin{center}
  $
  \begin{array}{c@{\hspace{0.1in}}c@{\hspace{0.1in}}c@{\hspace{0.1in}}c}
  &
  \includegraphics[scale=0.17]{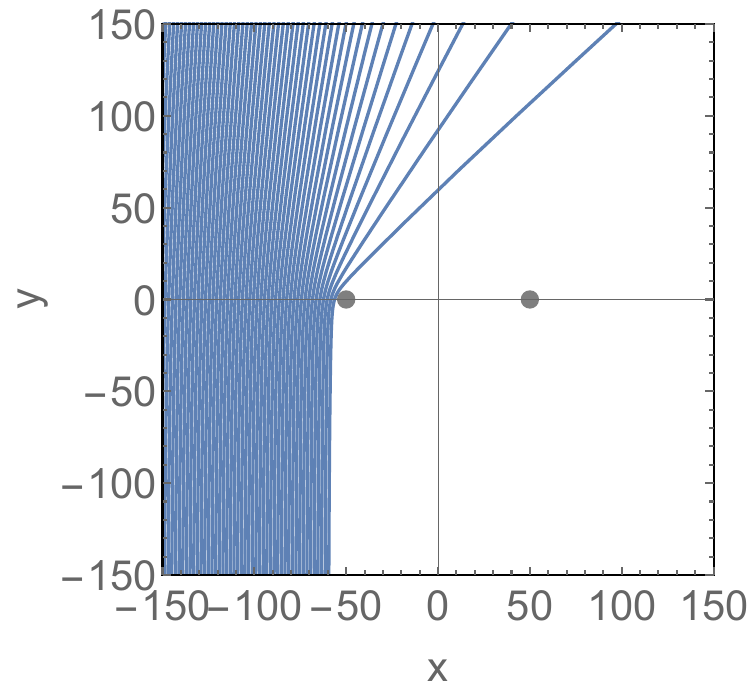} &
  \includegraphics[scale=0.17]{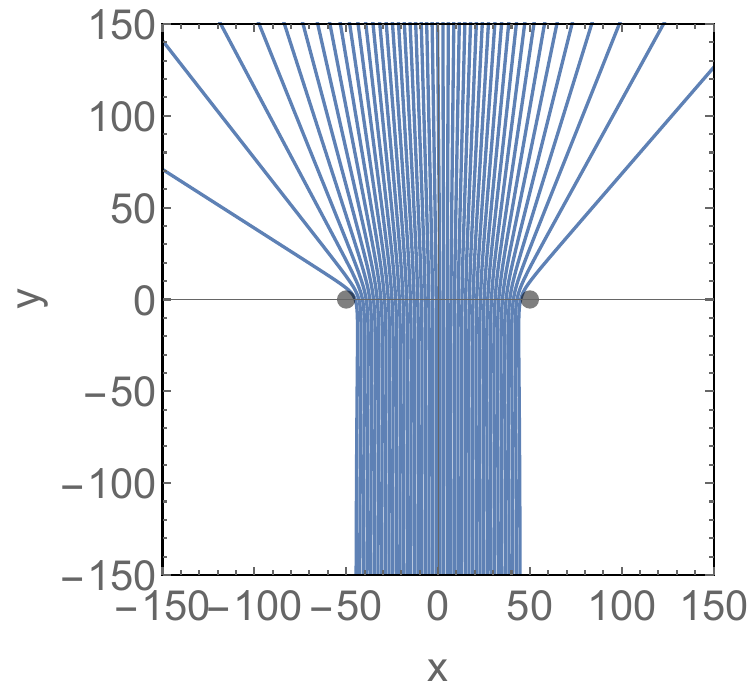} &
  \includegraphics[scale=0.17]{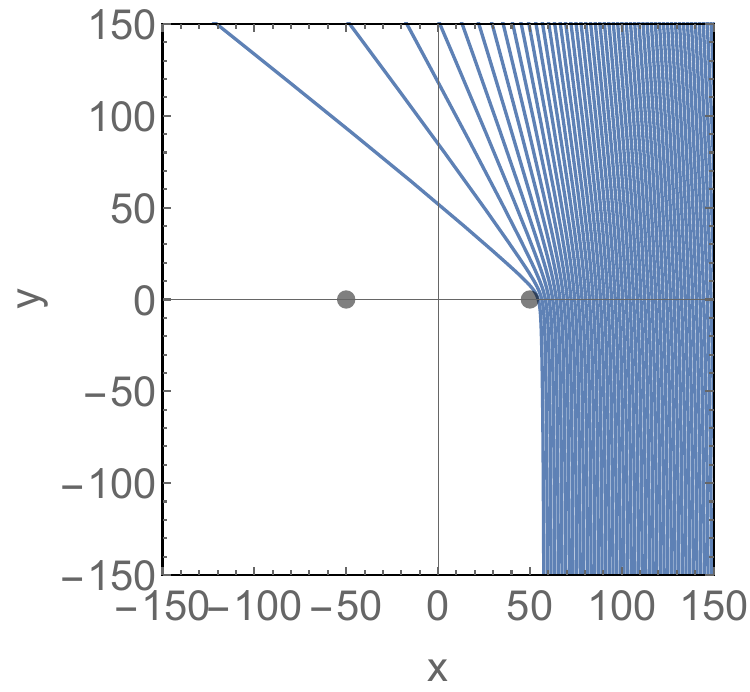}\\
&  &\vspace{0.2in}\mbox{\bf (a)}   &  \\
   &
  \includegraphics[scale=0.17]{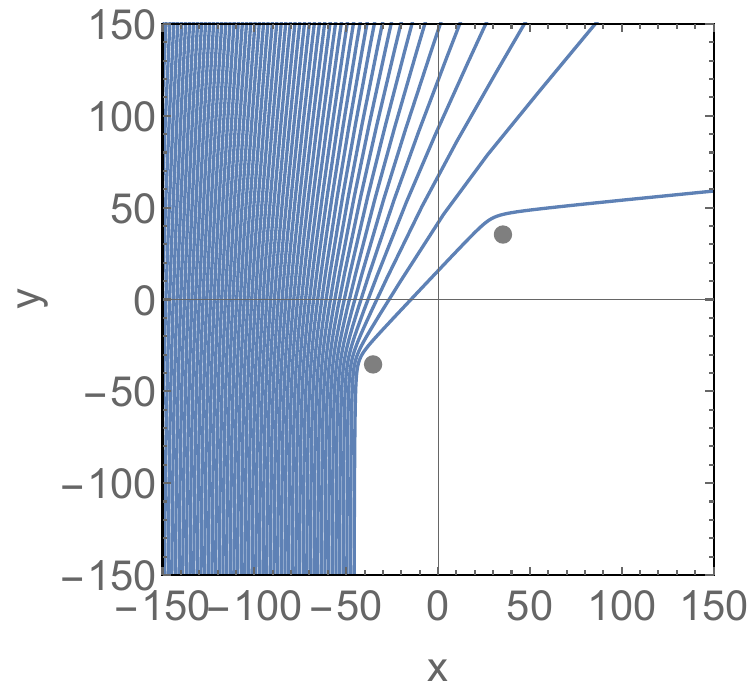} &
  \includegraphics[scale=0.17]{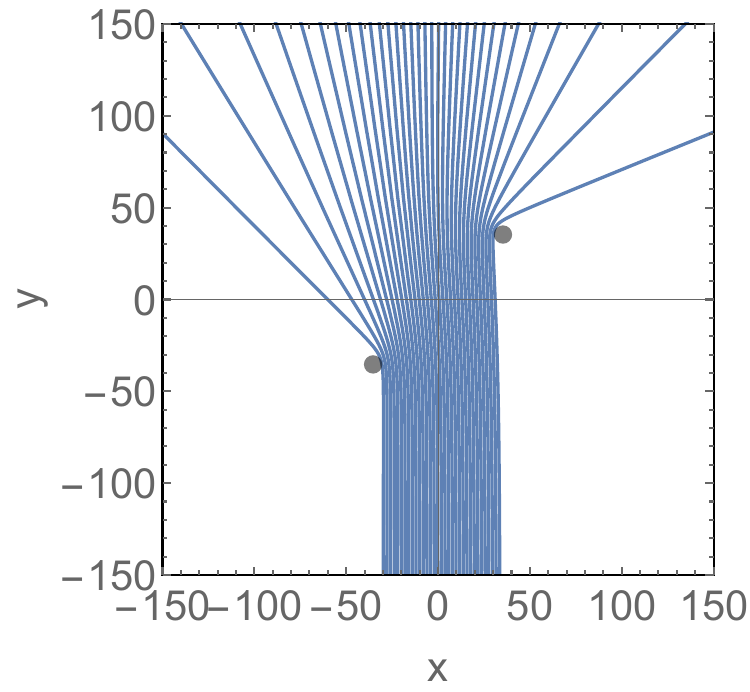} &
  \includegraphics[scale=0.17]{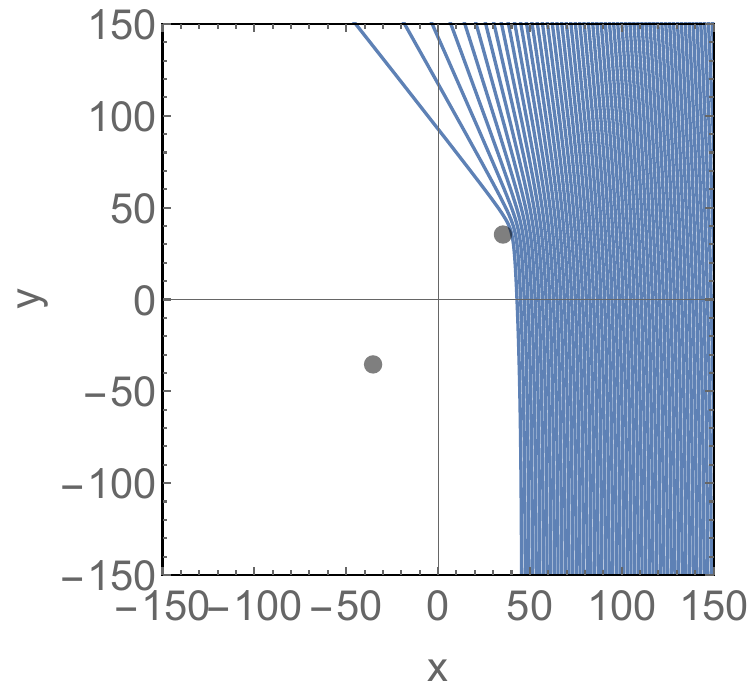}\\
  &  &\vspace{0.2in}\mbox{\bf (b)}   &  \\
  \end{array}
  $
  $
  \begin{array}{c@{\hspace{0.08in}}c@{\hspace{0.08in}}c@{\hspace{0.08in}}c@{\hspace{0.08in}}c}
   &
  \includegraphics[scale=0.15]{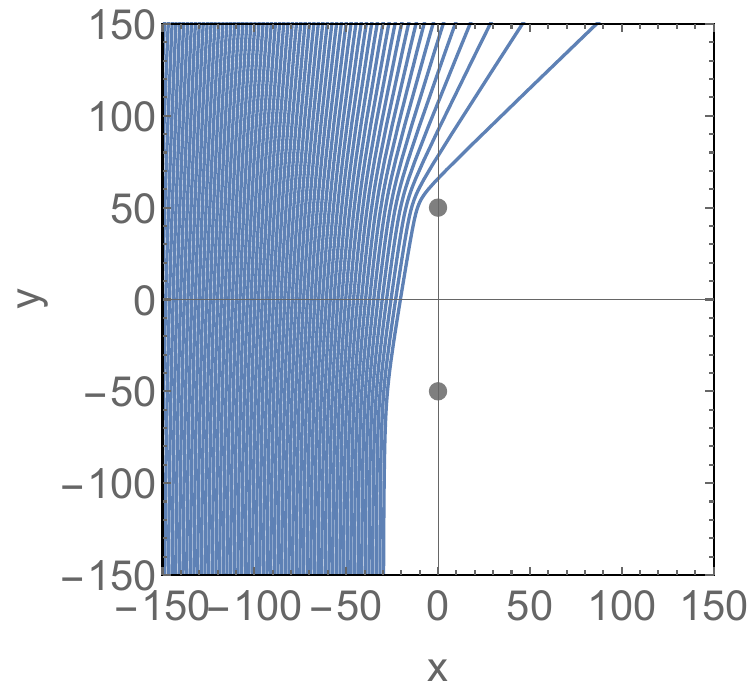} &
  \includegraphics[scale=0.15]{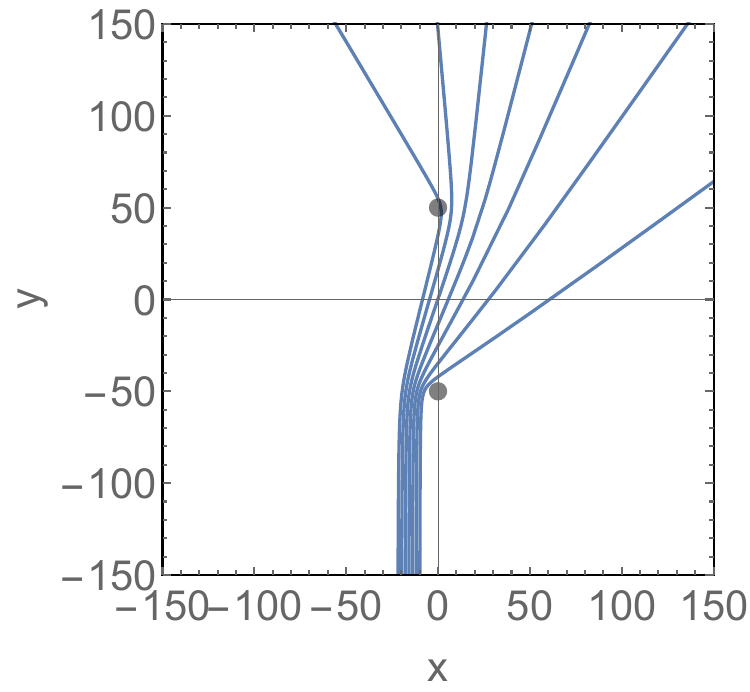} &
  \includegraphics[scale=0.15]{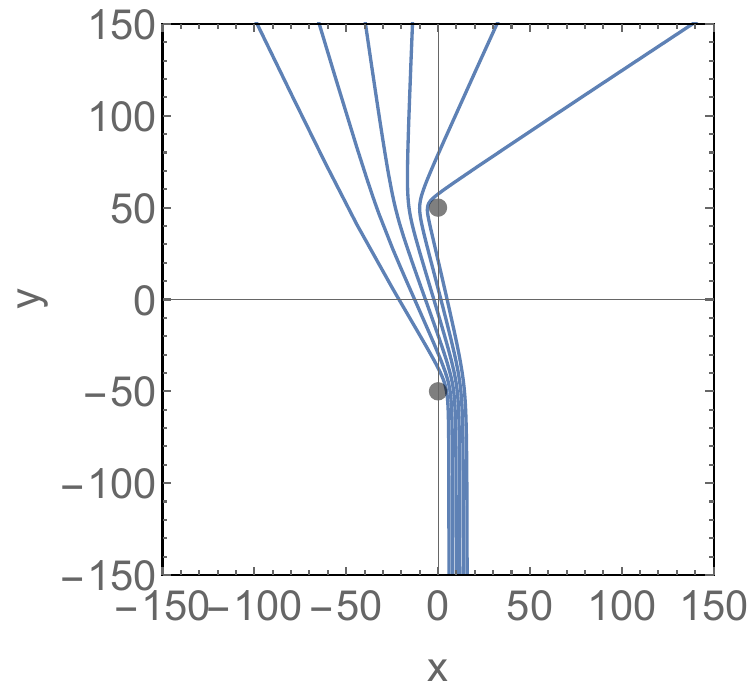} &
  \includegraphics[scale=0.15]{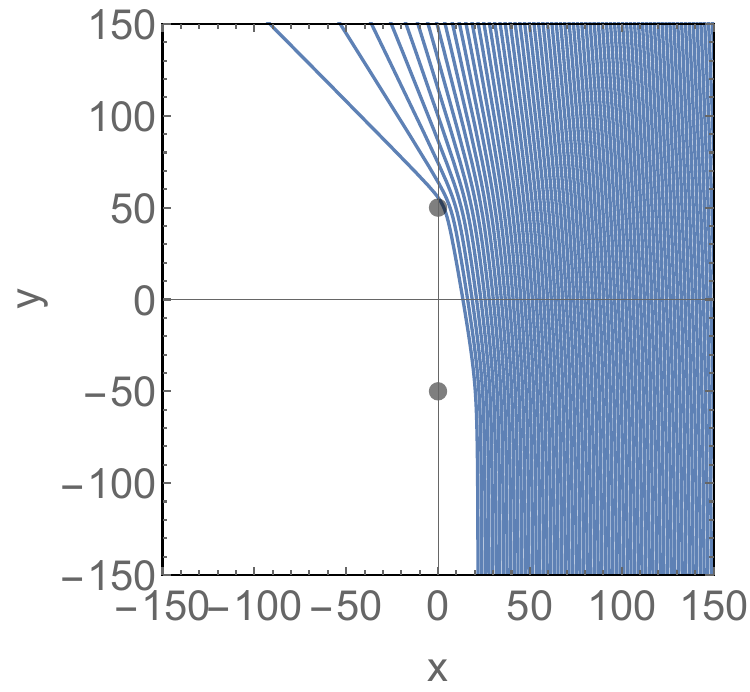} \;\;\;\;\;\;\;\;\;\;\;\;\;\;\;\;\;\;\\
  &  &  &\vspace{0.2in}\mbox{\bf (c)} \;\;\;\;\;\;\;\;\;\;\;\;\;\;\; \\
  \end{array}
  $
  \caption{The diffraction paths of incoming parallel wave from infinity bends according to the angle between in relative position vector of the two black holes and the incoming wave vector, (a) $\pi/2$, (b) $\pi/4$, and (c)$\pm\pi$.}
    \label{fig:three parts}
\end{center}
\end{figure}

The incoming parallel waves bend their paths in vicinity of the binary black hole (Fig. \ref{fig:three parts}), showing similar but different diffraction phenomena from that of light. We choose the following initial parameters, the initial mass of the two black holes $m_1 = m_2 = 1$, the distance between the two black holes $b = 100$, the orbital period $T_{BH} = 4443$ and the angular velocity $\omega_{BH} = \sqrt{2}/1000$. The incoming monochromatic plane wave is assumed to come from infinity. The two black holes are far separated from each other, the fully absorbed waves by the two black holes occupy a negligible small portion of the incoming wave. Fig. \ref{fig:three parts} shows the diffraction of long waves when the relative position vector of the binary black hole $\boldsymbol{r}_{12} = {\boldsymbol{x}_1-\boldsymbol{x}_2}$ rotates to an angle of $\pi/2$ ( in Fig. \ref{fig:three parts} (a)), $\pi/4$ (in Fig. \ref{fig:three parts} (b)) and $\pm\pi$ ( in Fig. \ref{fig:three parts} (c)) from the incoming wave vector $\boldsymbol{k}$. The unabsorbed incoming waves on the left (right) handed side bend to the right (left) hand side when they propagate in the perpendicular direction of $\boldsymbol{r}_{12}$ in Fig. \ref{fig:three parts} (a). The incoming wave passing through the middle zone between two black holes spreads into a symmetric fan shaped output beams in Fig. \ref{fig:three parts} (a). When the binary black hole rotates by an angle of $\pi/4$, the diffraction of long wave on the left (right) handed side of the black hole No. 1 (No. 2) is highly enhanced (suppressed). The output beams out of the middle zone demonstrates a highly asymmetric diffraction pattern in Fig. \ref{fig:three parts} (b). When the relative position vector of the two black holes $\boldsymbol{r}_{12}$ rotate to the parallel direction to the incoming wave vector $\boldsymbol{k}$, beside the suppressed diffraction on both the left-handed and the right handed side of the two black holes, the incoming waves passing through the middle zone of the two black holes propagates forwardly in an $"S"$-shaped trajectory in Fig. \ref{fig:three parts} (c).

\begin{figure}[ht]
\begin{center}
  $
  \begin{array}{c@{\hspace{0.1in}}c@{\hspace{0.1in}}c@{\hspace{0.1in}}c}
  \includegraphics[scale=0.2]{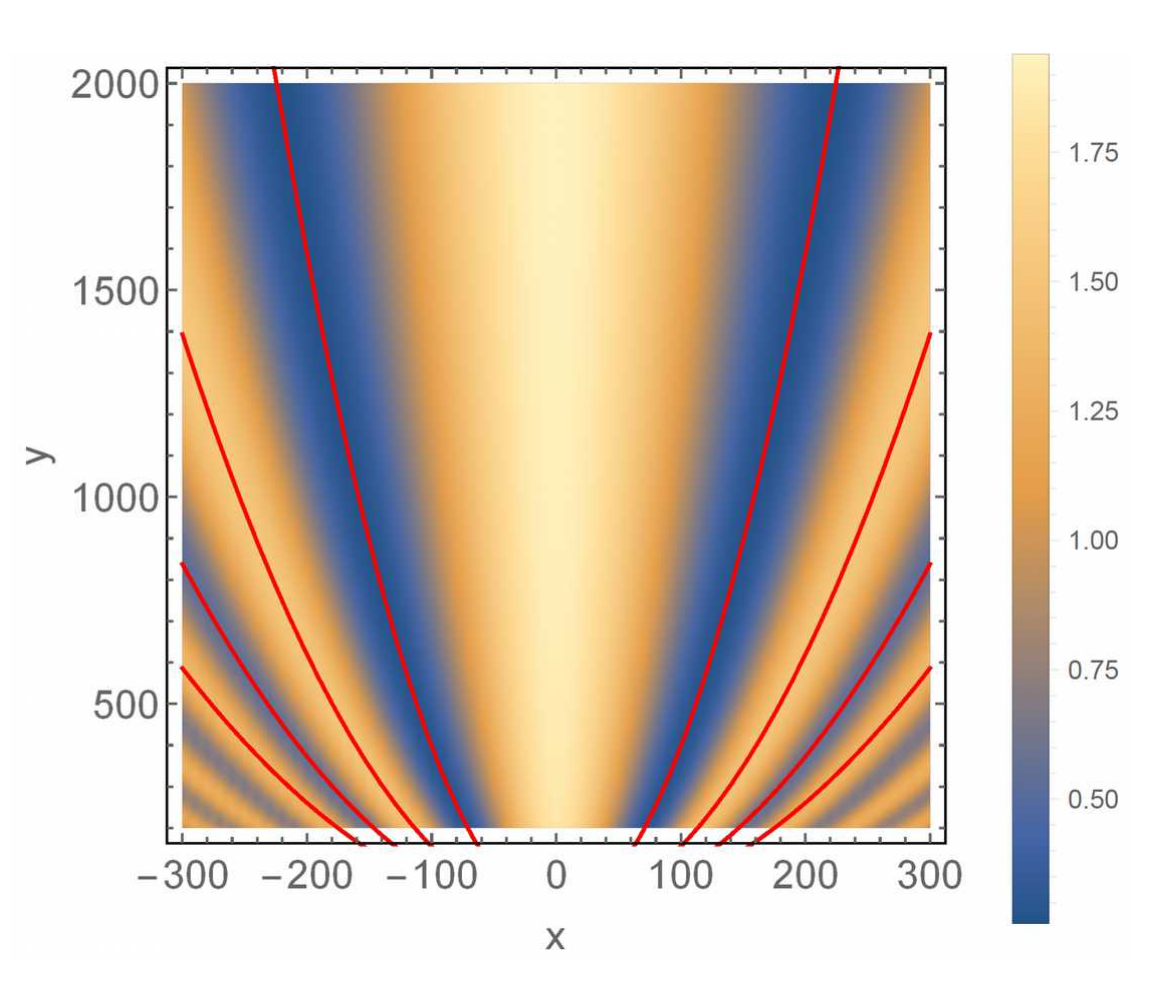} &
  \includegraphics[scale=0.2]{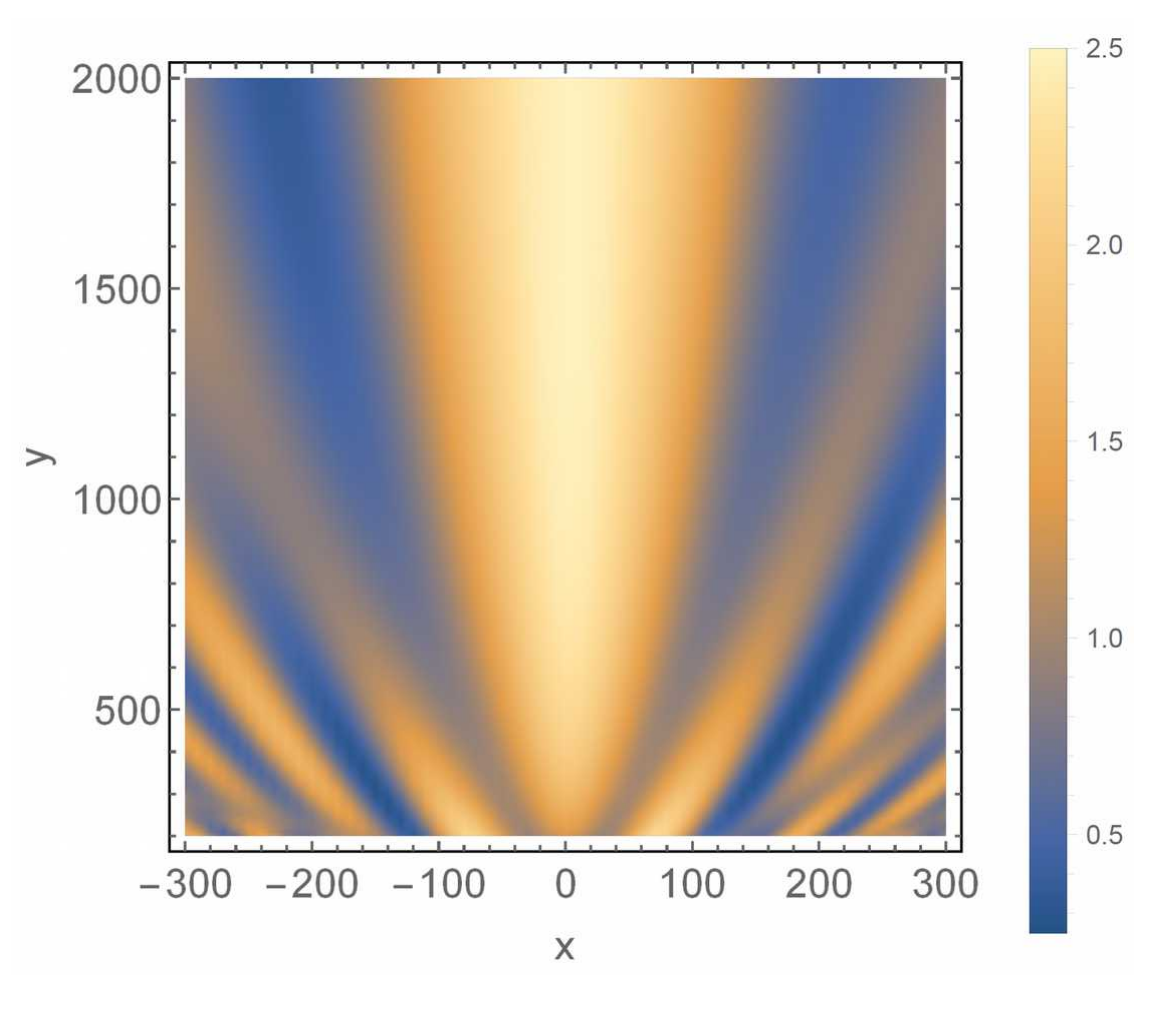} \\
  \mbox{\bf (a)} & \mbox{\bf (b)} \\
  \includegraphics[scale=0.3]{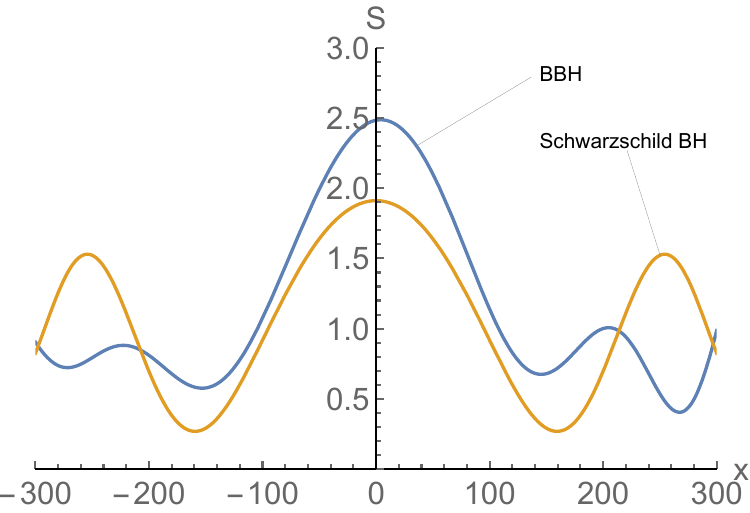}&
  \includegraphics[scale=0.2]{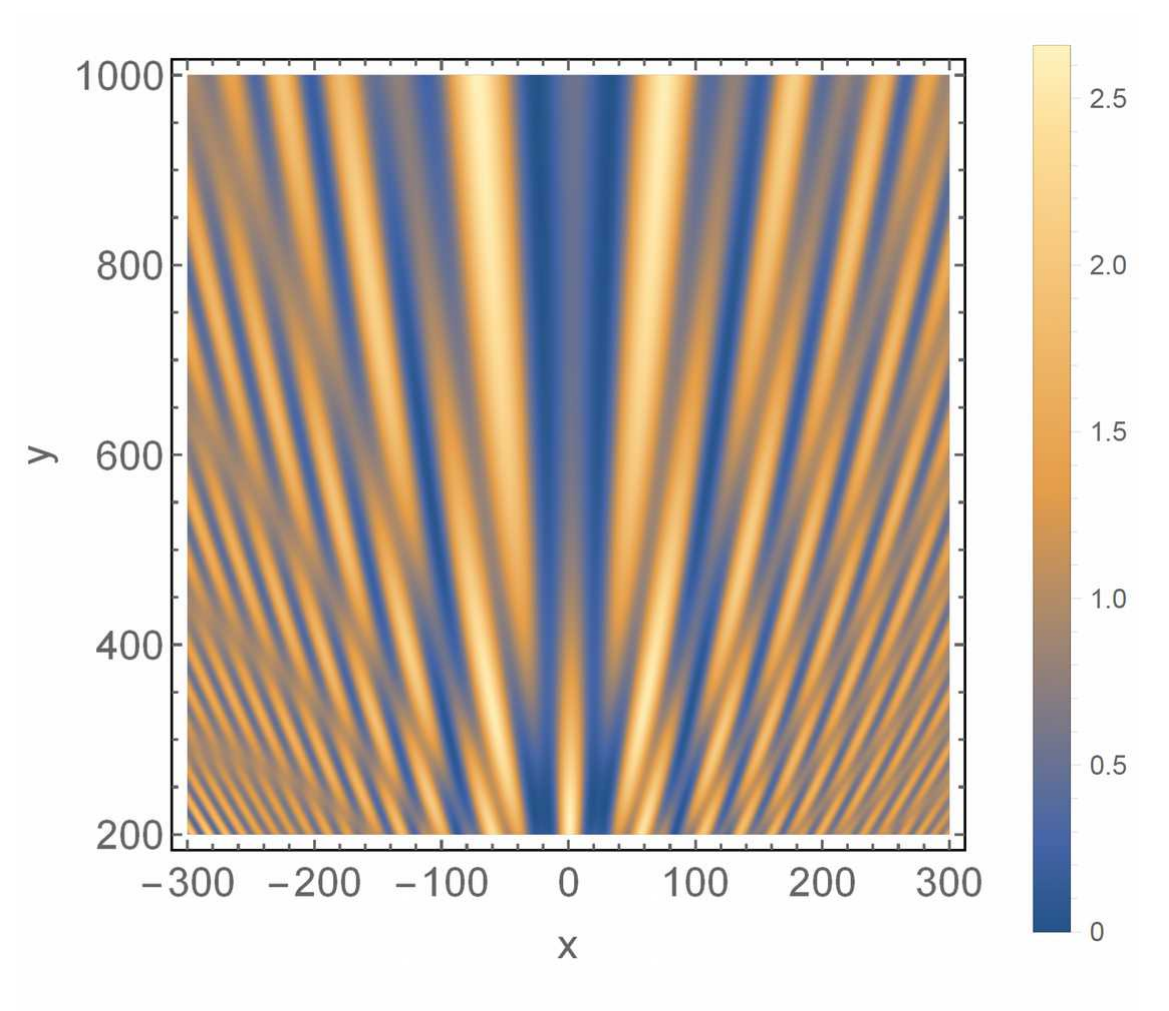} \\
  \mbox{\bf (c)} & \mbox{\bf (d)} \\
  \end{array}
  $
  \caption{(a) Interference of $\lambda=40$ monochromatic plane wave passing through a Schwarzschild black hole with $m=2$, (b) Interference of $\lambda=40$ monochromatic plane wave passing through a binary black hole, (c) Comparison of light intensity between the binary black hole and the Schwarzschild black hole, at $y=1000$, (d) the interference of the monochromatic plane wave with $\lambda=10$ passing through the binary black hole.}
     \label{fig:interference}
\end{center}
\end{figure}

When the incoming parallel waves pass the left-handed, the right-handed and the middle zone of the two black holes simultaneously, they intersect with one another, resulting in interference pattern in the outgoing zone (as shown in Fig. \ref{fig:interference}). The incoming wave is the superposition of three monochromatic waves within the left handed zone, the middle zone and the right handed zone correspondingly, in analogy with slits of the blocking board of double slits experiments,
\begin{equation}
S =\left|\sqrt{S_l}e^{i\omega t_l}+\sqrt{S_m}e^{i\omega t_m}+\sqrt{S_r}e^{i\omega t_r}\right|^2.
\end{equation}
The projection screen is placed at infinity( here we choose $y =1000$). The $i$-th geodesic path intersects the projection screen at $(t_i,x_i)$. The wave intensity is defined the density of the geodesic: $S_i = 1/[(x_{i+1}-x_{i-1})cos\theta_i]$, where $\theta_i$ is the angle between the geodesic path and the $y$-axis at the intersection point. The local time $t_{\alpha}$, $ (\alpha = l, m, r)$ is determined by the spacetime of the binary black hole. The wavelength of the incoming wave is chosen as comparable with the Schwarzschild radius of a black hole in order to detect an apparat interference pattern. For a small black hole with a radius of $\lambda = 3*10^3 m$, the frequency of electromagnetic wave is $\omega = 10^5 Hz$. The spatial distribution of amplitude of interference waves in x-y plane is numerically computed by moving the location of the projection screen continuously, as showed in Fig. \ref{fig:interference}, the bright (dark) region has higher (lower) intensity. This interference pattern on a light-like hypersurfaces is derived from the same set of geodesics. Fig. \ref{fig:interference} (a) shows the interference patterns of long waves with $\lambda = 40$ (dimensionless) passing through one Schwarzschild black hole with $m = 2$ at $x = 0$. The brightest wave packet propagates along the $y-$axis at $x = 0$, accompanied by two second brightest peaks on the left handed and the right handed region. When the same set of long waves pass through a binary black hole (Fig. \ref{fig:interference} (b)), the brightest wave packet is widened but still travels along the perpendicular line to the bonding vector that connects the two black holes $\boldsymbol{r}_{12}$. The central peak is generated by the superposition of waves with the same phase, since the incoming waves from both the left handed side and the right handed side travel over the same distance to reach the perpendicular line. The dimensionless relative wave intensity approaches to 2 at infinity. The second brightest peaks are highly suppressed, as showed in Fig. \ref{fig:interference} (b). The convergence effect of the binary black hole is stronger than single black hole. Fig. \ref{fig:interference} (c) showed the spatial distribution of intensity, the central peak in the outgoing zone of the binary black hole decays faster than that of single black hole.

The maximal peak of the superposition waves after passing single black hole is always along the middle line despite of wavelengths, i.e., the $y$-axis at $x = 0$ (Fig. \ref{fig:interference} (a)). However the interference pattern in the outgoing zone of binary black hole has strong dependence on the wavelength of the incoming waves. Fig. \ref{fig:interference} (d) showed the interference pattern of incoming waves with a relatively short wavelength $\lambda = 10$. In the near zone of the binary black hole, the maximal peak is still located along the middle line. The brightest peak decays as the observer travels away from the binary black hole, so does the other bright peaks. On the contrary, the intensity of the dark peaks grow to become the bright peaks in the far zone of the binary black hole. Therefore the intensity of superposition waves along the perpendicular line to the bonding line between the two black holes is only strengthened in the near zone, and is reduced to zero in the far zone of the binary black hole.

\begin{figure}[ht]
\begin{center}
  $
  \begin{array}{c@{\hspace{0.1in}}c@{\hspace{0.1in}}c@{\hspace{0.1in}}c}
  \includegraphics[scale=0.2]{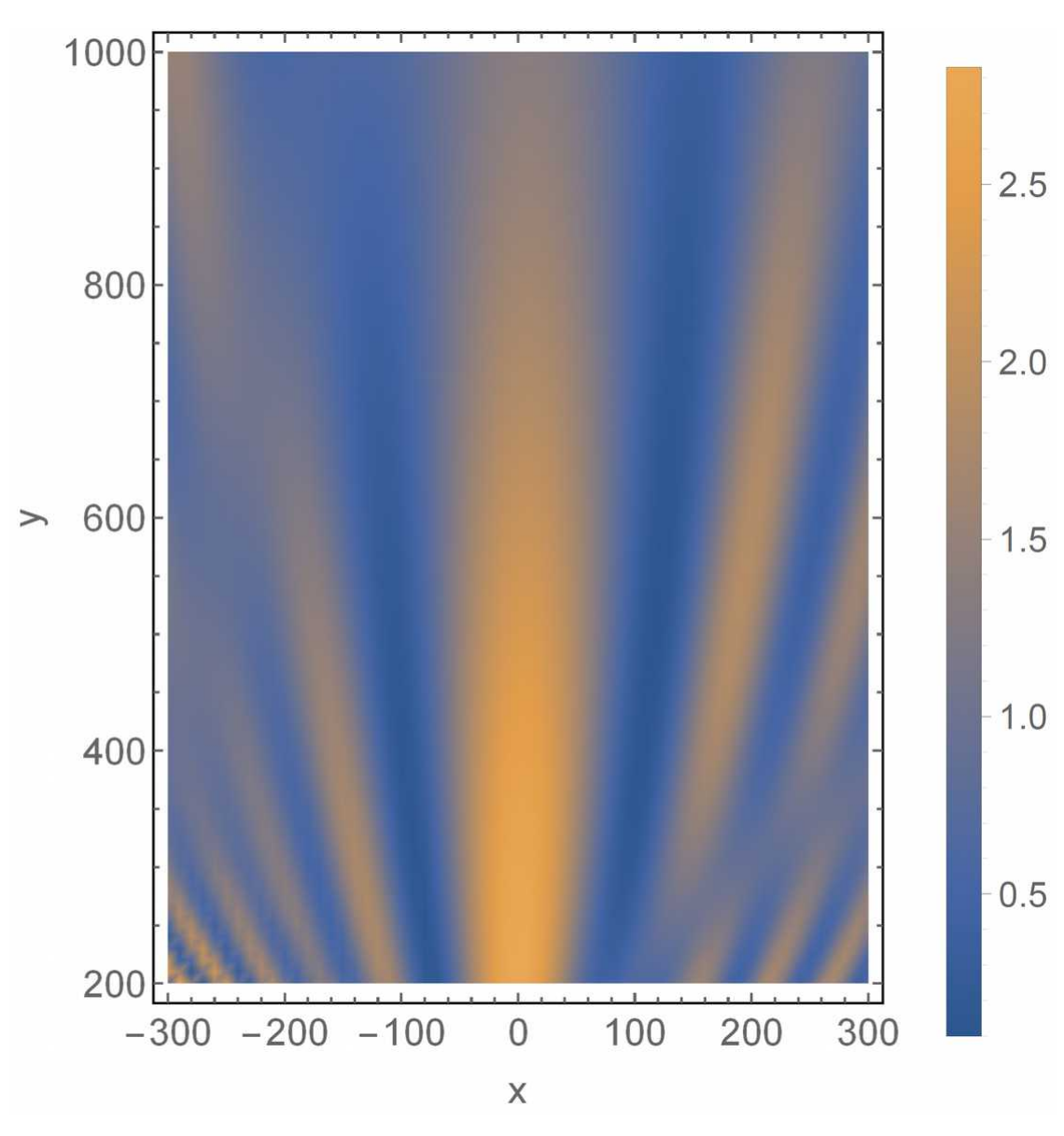} &
  \includegraphics[scale=0.2]{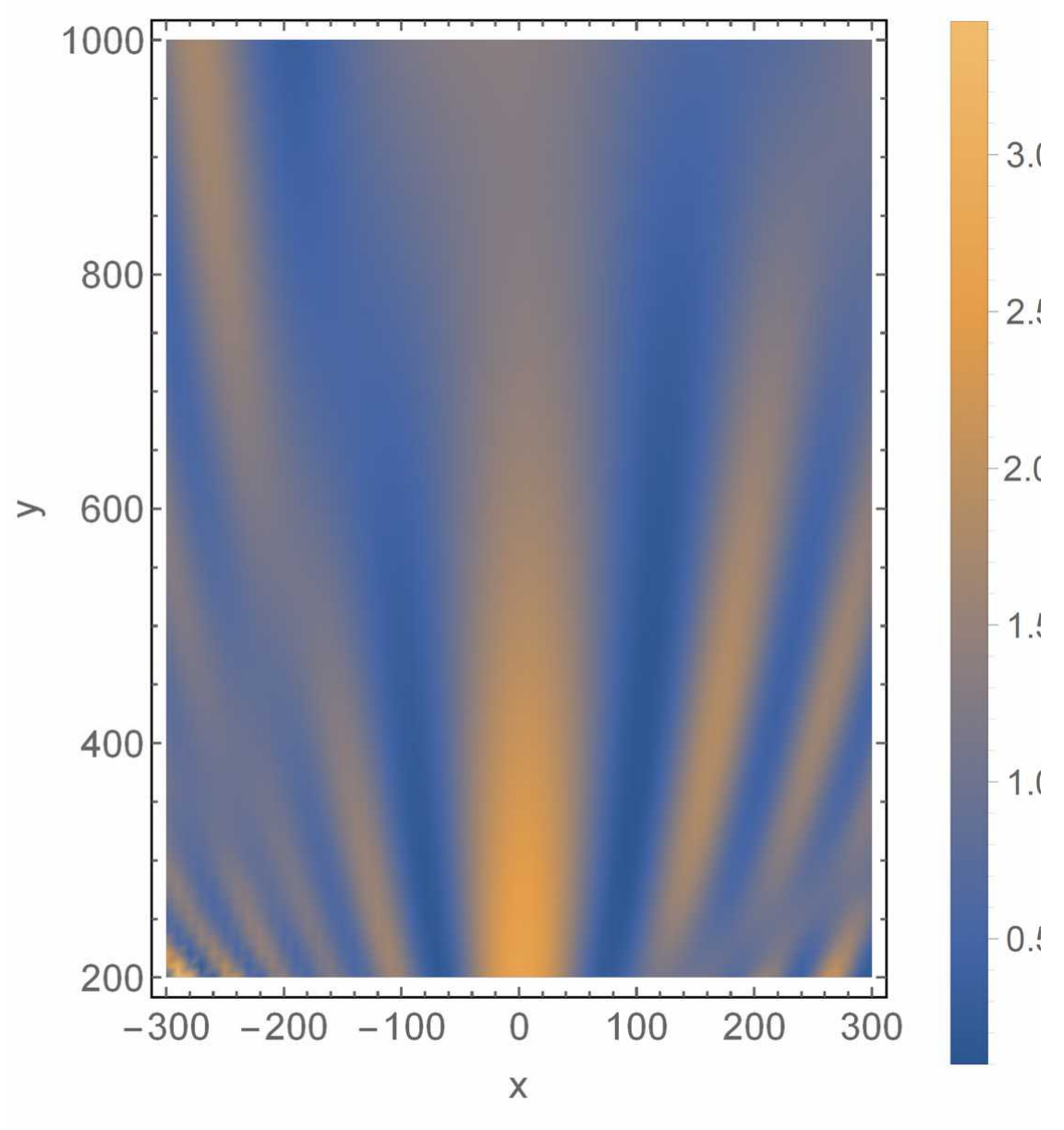} \\
  \mbox{\bf (a) $50^{\circ}$} & \mbox{\bf (b) $60^{\circ}$} \\
  \includegraphics[scale=0.2]{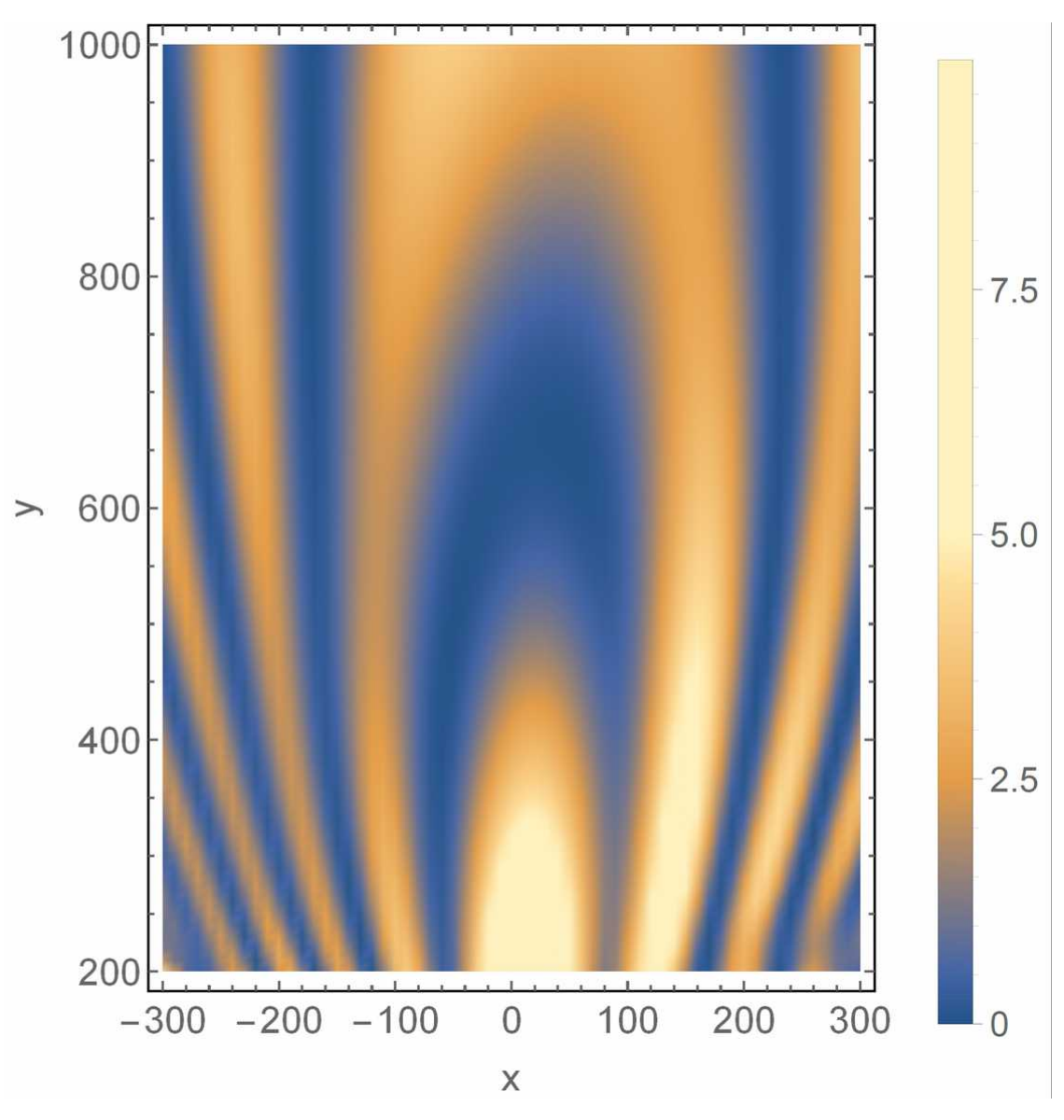} &
  \includegraphics[scale=0.2]{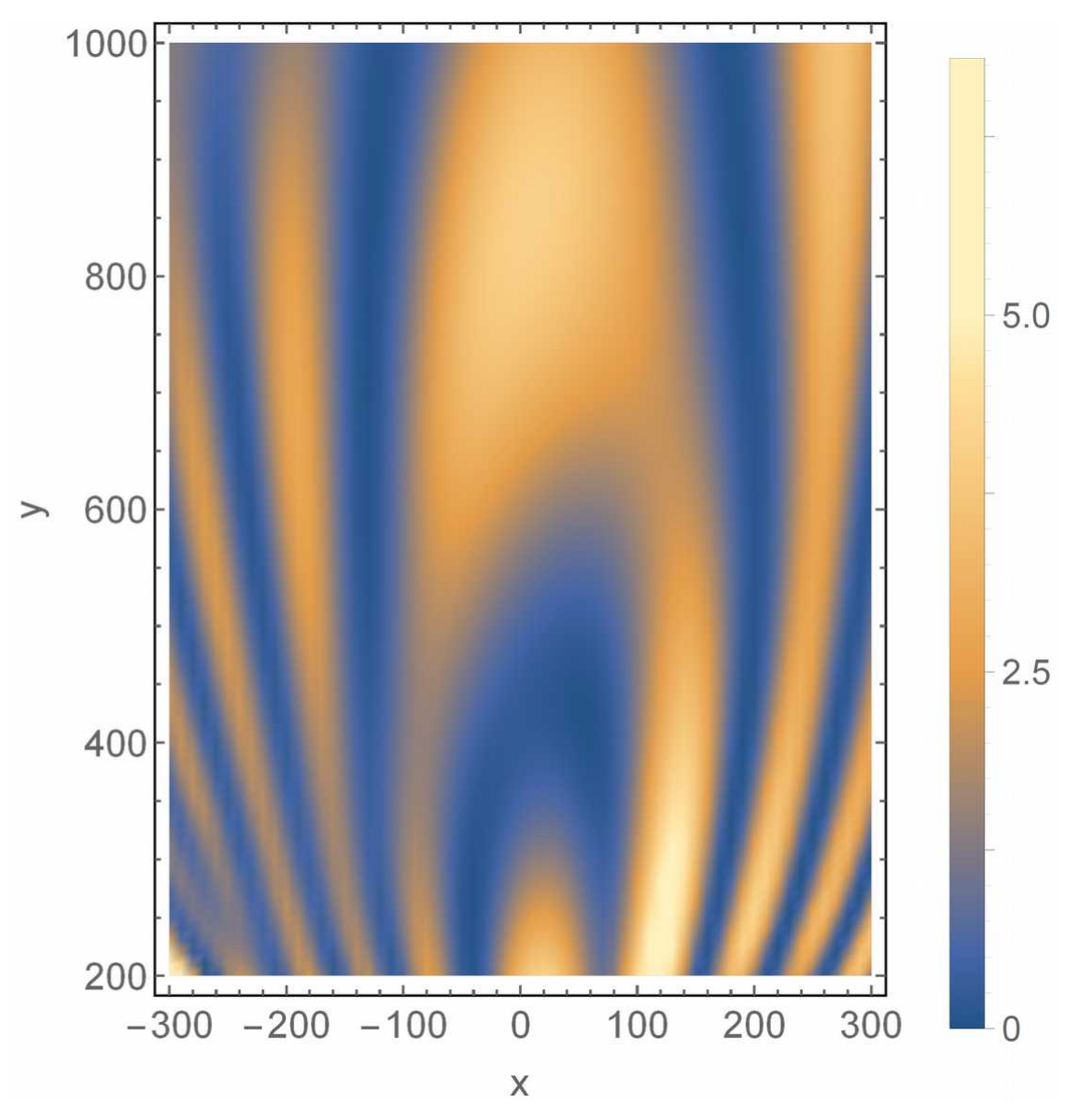} \\
  \mbox{\bf (c) $70^{\circ}$} & \mbox{\bf (d) $80^{\circ}$} \\
  \end{array}
  $
  \caption{The interference patterns when the black hole rotates to angle of $50^{\circ}$, $60^{\circ}$, $70^{\circ}$ and $80^{\circ}$.}
    \label{fig:Shine}
\end{center}
\end{figure}

The two black holes are located along $x$-axis in Fig. \ref{fig:interference}, i.e., the angle between the relative position vector $\boldsymbol{r}_{12}$ and $x$-axis is $0^{\circ}$. The interference pattern shows different spatial distribution when the bonding vector $\boldsymbol{r}_{12}$ rotates to $50^{\circ}$, $60^{\circ}$, $70^{\circ}$ and $80^{\circ}$, as showed in Fig. \ref{fig:Shine}. When the bonding vector is further tilled to $50^{\circ}$ and $60^{\circ}$, the outgoing waves split into straight rays distributed in bright and dark pattern alternatively. These straight rays decays in a straight way to infinity (Fig. \ref{fig:Shine} (a-b)). When the bonding vector is further tilled to $70^{\circ}$, the bright rays generates by interference bend into parabolic arcs that bridges the two black holes (Fig. \ref{fig:Shine} (c)). A further rotation of $\boldsymbol{r}_{12}$ to $80^{\circ}$ draws the bright ray bridge closer to the binary black hole. In the mean time, the intensity of the outgoing wave becomes more than ten times stronger  (Fig. \ref{fig:Shine} (d)).

\begin{figure}[ht]
\begin{center}
 \includegraphics[scale = 0.7]{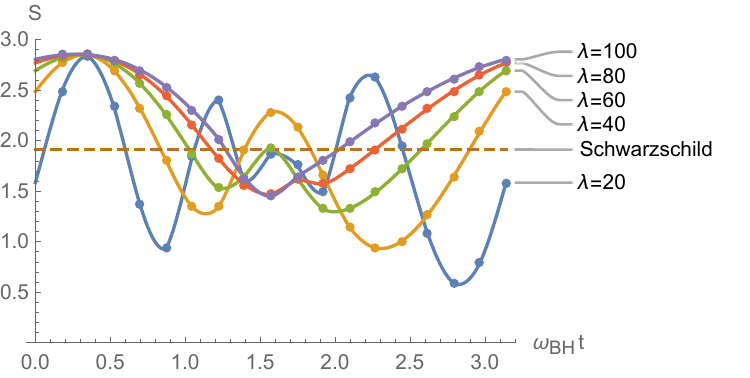}
  \caption{ The evolution of interference intensity at different wavelengths over time at point $(0,1000)$. }
    \label{fig:wavelengths}
\end{center}
\end{figure}

The intensity of interference pattern changes with respect to different wavelengths and phase differences of the incoming waves. For single black hole, the intensity of interference pattern is independent of wavelength ( as shown by the dash line in Fig. \ref{fig:wavelengths}). At a point $(x,y)=(0,1000)$ on the perpendicular axis to the bonding vector $\boldsymbol{r}_{12}$ of the binary black hole, the light intensity oscillates with different wavelengths, as shown in Fig. \ref{fig:wavelengths}. A minimal intensity locates at the $\theta = \omega_{H}t = \pi/2$ in Fig. \ref{fig:wavelengths}. The number of periodical oscillations increases when the wavelength is reduced (Fig. \ref{fig:wavelengths}).

\begin{figure}[ht]
\begin{center}
 $
 \begin{array}{c@{\hspace{0.1in}}c@{\hspace{0.1in}}c}
 \multicolumn{1}{l}{\mbox{}} & \multicolumn{1}{l}{\mbox{}} & \multicolumn{1}{l}{\mbox{}} \\
 \includegraphics[scale=0.14]{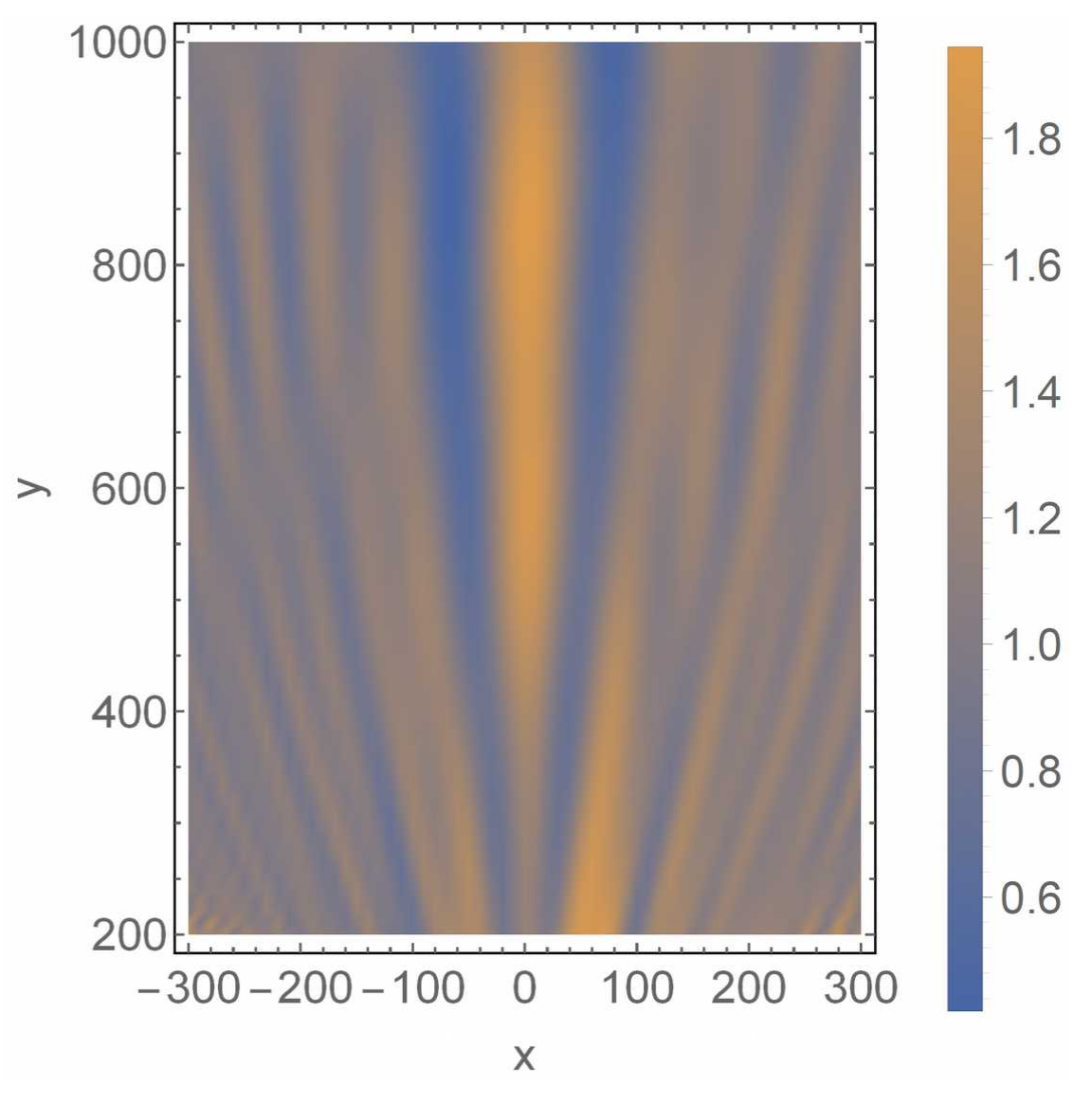} &
 \includegraphics[scale=0.14]{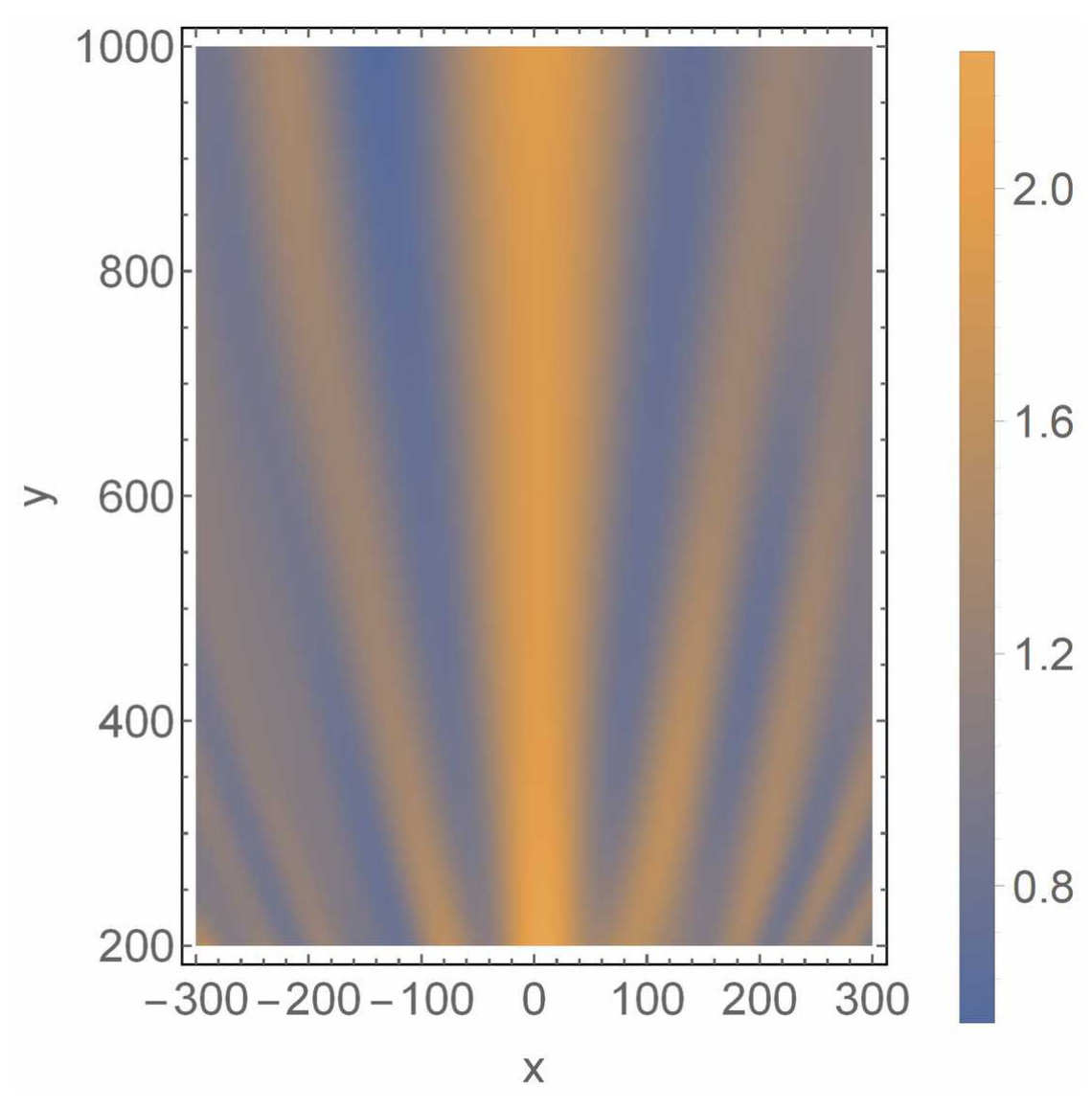} &
 \includegraphics[scale=0.14]{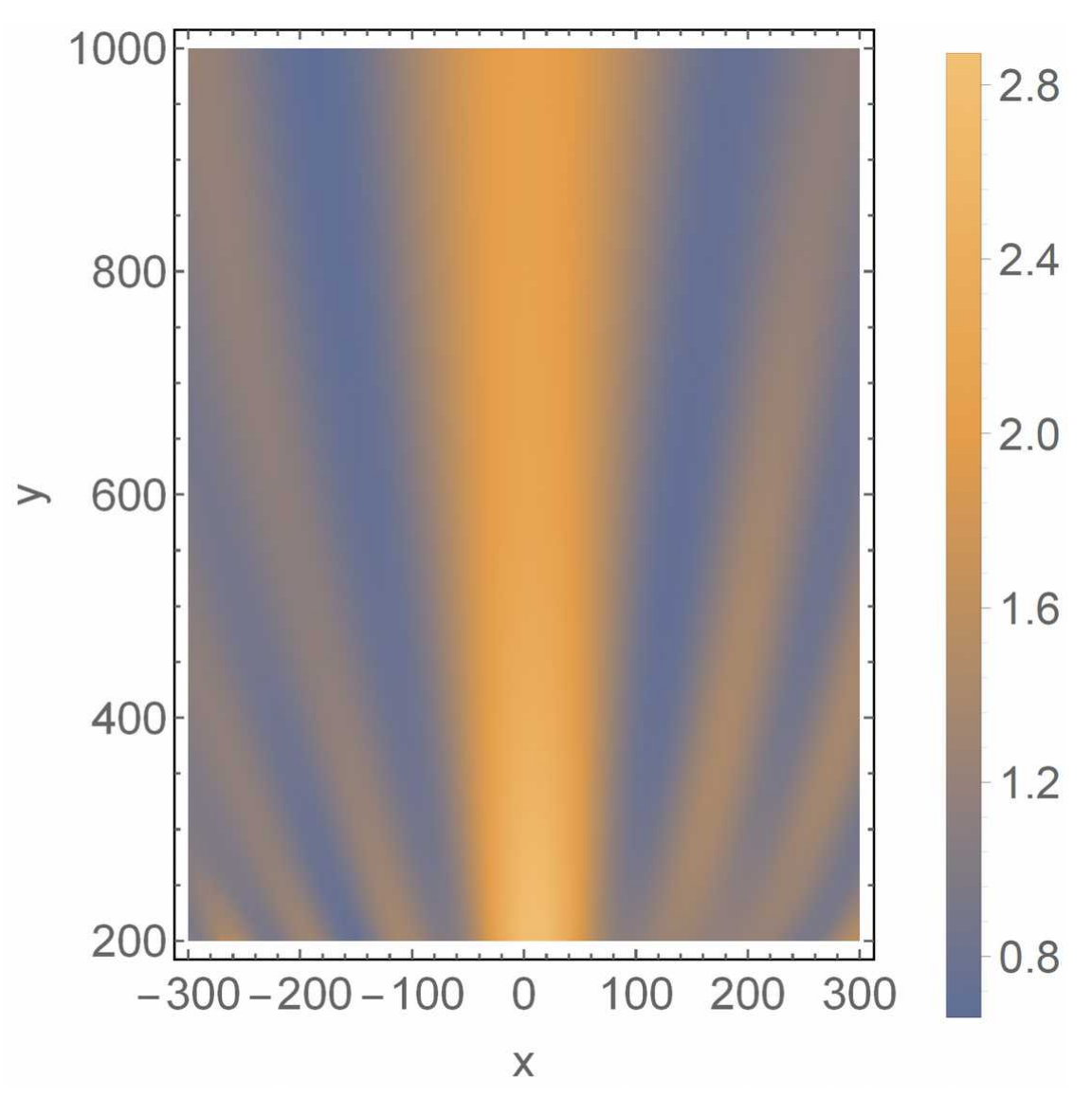}\\
 \mbox{\bf (a)} & \mbox{\bf (b)} & \mbox{\bf (c)}
 \end{array}
 $
  \caption{The average interference intensity of light in one orbital period. (a) $\lambda=20$ (b) $\lambda = 40$ (c) $\lambda = 60$.}
    \label{fig:average}
\end{center}
\end{figure}

\begin{figure}[ht]
\begin{center}
 \includegraphics[scale=0.6]{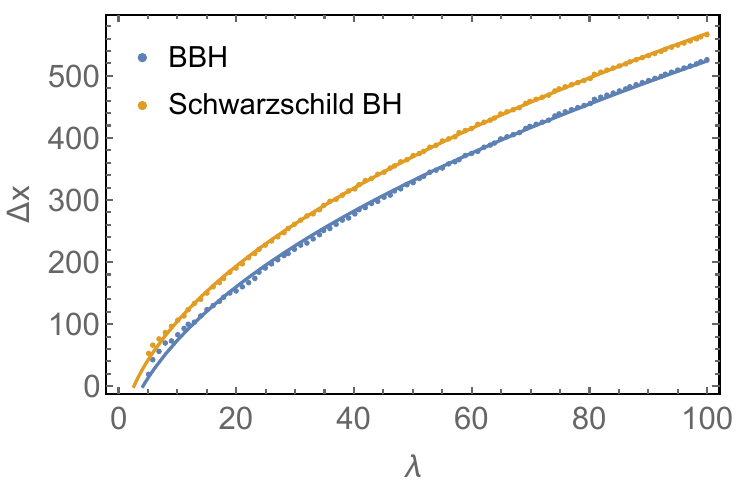}
  \caption{The width of the central bright fringe as a function of the wavelength of the incident light. }
    \label{fig:fit}
\end{center}
\end{figure}

Interfering light intensity is a function of space-time coordinates, $S(t,x,y)$. We average the light intensity of each point within a binary black hole orbit period,
\begin{equation}\label{average}
\overline{S}(x,y)=\frac{1}{n}\sum_{i=1}^n{S\left( t_i,x,y \right)}.
\end{equation}
The average light intensity always shows the maximum value along the perpendicular axis to $\boldsymbol{r}_{12}$. The central bright fringe sandwiched in between two dark fringes is most obvious for short wavelength (Fig. \ref{fig:average}). The width of the central bright fringe widens when the wavelength increases, from $\lambda = 20$, to $\lambda = 40$ and $\lambda = 60$ (Fig. \ref{fig:average} (a-b-c)). The width of the central bright fringe generated by the lens effect of two black holes is wider than that of single Schwarzschild black hole, which approximately obeys $\Delta x\propto \sqrt{x}$. The width of the central bright fringe of the binary black hole increases slowly as the observer moves away from the binary black hole, as shown in Fig. \ref{fig:average}. On the detection screen at $y=1000$, the width of the central bright fringe of the average wave intensity in Fig. \ref{fig:fit} is approximately fitted by $\Delta x\propto a \sqrt{\lambda} + b $, with fitting parameter $a=67.8$, $b=-110$ for single Schwarzschild black hole and $a=65.8$, $b=-134$ for the binary black holes (Fig. \ref{fig:fit}). The intensity of the central bright fringe increases with the increasing of incoming wavelength. The peak light intensities are $\overline{S}_{\mathrm{max}}=1.75, 1.82, 2.20, 2.43, 2.62$ with respect to $\lambda=20, 40, 60, 80, 100$.

\section{The Gravitational redshift around binary black hole}

\begin{figure}
  \centering
  \includegraphics[scale=0.40]{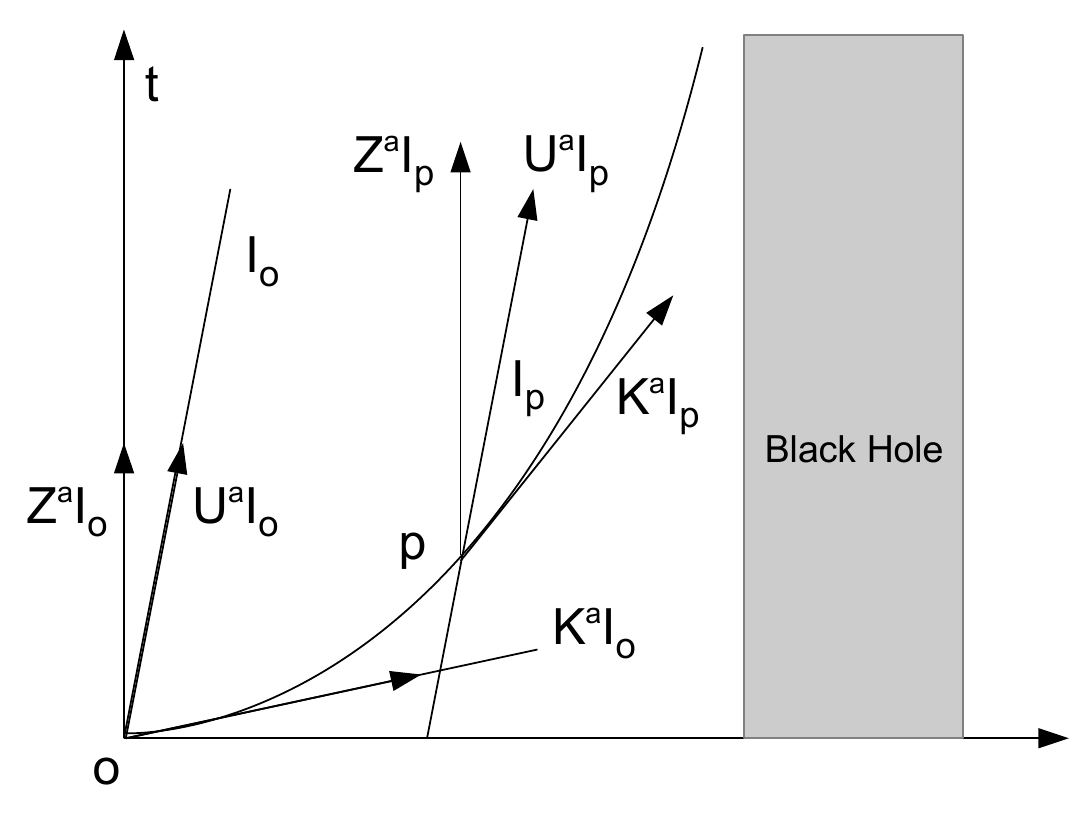}\\
  \caption{The gravitational redshift measured by observers of a stationary and moving black hole in a coordinate system.}\label{worldline}
\end{figure}

A traveling wave moves away from a moving black hole experienced gravitational redshift. To find the gravitational redshift of light around a moving black hole in an equivalent way, we establish a stationary coordinate system on a black hole, and place a series of moving observers in parallel to find the frequency of light detected by these observers. The world lines of these moving observers are parallel to one another,  as shown in Fig. \ref{worldline}. The frequency of a light traveling from a point $\mathrm{o}$ to a point $\mathrm{p}$ in curved spacetime is
\begin{equation}\label{eq1-1}
  \omega =-K^aU^bg_{ab},
\end{equation}
where $K^a$ is the wave vector of light, $U^b$ is the 4-dimensional velocity of the observer, and $g_{ab}$ is the metric of the point. In Fig. \ref{worldline}, the stationary observers $\left. Z^a \right|_{\mathrm{o}}$ and $\left. Z^a \right|_{\mathrm{p}}$ measure the frequency $\omega$ and $\omega'$ at points $\mathrm{o}$ and $\mathrm{p}$. The moving observers $\left. U^a \right|_{\mathrm{o}}$ and $\left. U^a \right|_{\mathrm{p}}$ measures the frequency $\omega _{\mathrm{o}}$ and $\omega _{\mathrm{p}}$. The world line $l_o$ of a moving observer is parallel to $l_p$ in Fig. \ref{worldline}. The wave vector of light along the world line of the point $o$ is $\left. K^a \right|_{\mathrm{o}}=\left(\omega , \vec{k} \right) $, where $k^2=\omega^2$. The velocity of the moving observer is $\left. U^a \right|_{\mathrm{o}}=\left( \gamma , \gamma \vec{u} \right) $, where $\gamma =1/\sqrt{1-{u}^2}$. The frequency of light measured by the moving reads
\begin{equation}\label{eq1-2}
\omega _{\mathrm{o}}=\gamma \omega \left( 1-u\cos \theta \right),
\end{equation}
where $\theta$ is the angle between $\vec{u}$ and $\vec{k}$. Eq. (\ref{eq1-2}) indicates the Doppler effect in flat spacetime.

The wave vector $\left. K^a \right|_{\mathrm{p}}$ at point $p$ is derived from the geodesic equation of a moving observer with a velocity $\left. U^a \right|_{\mathrm{p}}$,
\begin{equation}\label{eq1-3}
\begin{split}
\left. U^a \right|_{\mathrm{p}} =\gamma'\left. Z^a \right|_{\mathrm{p}}+\gamma' u'^a
=\left( \gamma'/\sqrt{-g_{00}}, \gamma'\vec{u}'\right),
\end{split}
\end{equation}
where $\gamma'=1/\sqrt{1-u'^au'_a}$, and $\left. U^a \right|_{\mathrm{p}}\left. U_a \right|_{\mathrm{p}}=-1$. The world line of the two observers, $l_{\mathrm{o}}$ and $l_{\mathrm{p}}$ are parallel, thus
\begin{eqnarray}\label{eq1-5}
    u^i = \left. \frac{U^i}{U^0} \right|_{\mathrm{p}}=\sqrt{-g_{00}}u'^i, \;\;\;
    \vec{u}' =\frac{\vec{u}}{\sqrt{-g_{00}}}.
\end{eqnarray}
Substituting the frequency Eq. (\ref{eq1-1}) into the Eq. ({eq1-5}) yields $\left. U^a \right|_{\mathrm{p}}$ and  $\omega _{\mathrm{p}}$. The gravitational redshift detected by moving observers is $\omega _{\mathrm{p}}/\omega _{\mathrm{o}}$.

\begin{figure}
  \centering
  \includegraphics[scale=0.6]{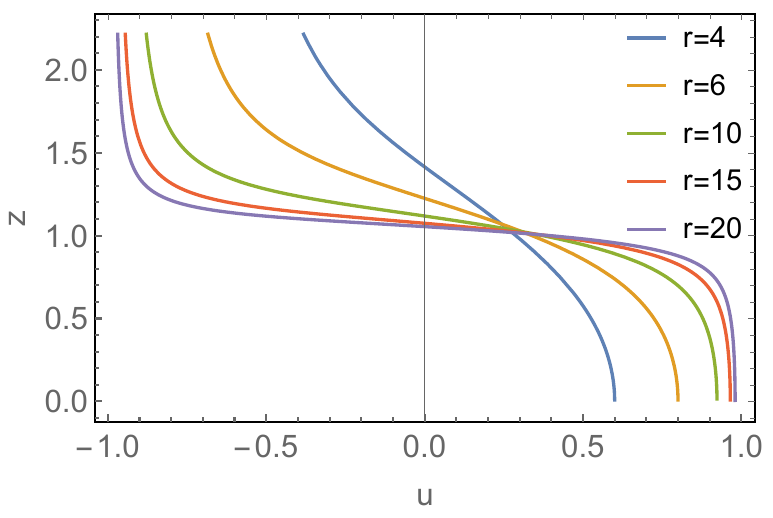}\\
  \caption{The velocity $u$ of the Schwarzschild black hole in the coordinate system is $r$ from the observer. The redshift $z=\omega_{p}/\omega_{o}$ measured by the observer at rest in a coordinate system.}\label{redshift0}
\end{figure}

We first apply this four dimensional equation of gravitational redshift for single Schwarzschild black hole moving at velocity of $\vec{u}=(-u,0,0)$ along $x$-axis. The light propagates in the positive direction of $x$-axis in Fig. \ref{worldline}. In the stationary coordinate system $\left( t,r,\theta ,\varphi \right) $ of the Schwarzschild black hole, the wave vector of the point $\mathrm{o}$ at infinity reads, $\left. K^a \right|_{\mathrm{o}}=\omega \left( 1 ,-1,0,0 \right) $. The moving observer at point $\mathrm{o}$ measured a frequency,
\begin{equation}\label{eq2-0}
  \omega _{\mathrm{o}}=\gamma \omega \left( 1-u \right),\;\;\;\gamma =\frac{1}{\sqrt{1-u^2}}.
\end{equation}
The distance between the point $\mathrm{p}$ and the black hole is $r$. The wave vector in Schwarzschild spacetime reads
\begin{equation}\label{eqkasch}
\left. K^a \right|_{\mathrm{p}}=\left( \frac{\omega}{1-\frac{2M}{r}},-\omega ,0,0 \right).
\end{equation}
In mind of the velocity Eq. (\ref{eq1-5}), the velocity of point $\mathrm{p}$ is
\begin{eqnarray}\label{eq2-1}
    u' = \frac{-u}{\sqrt{1-\frac{2M}{r}}},\;\;\;
   \gamma' = \frac{1}{\sqrt{1-\frac{u^2}{\left( 1-\frac{2M}{r} \right) ^2}}}.
\end{eqnarray}
Substituting the four dimensional velocity equation,
\begin{equation}\label{eq2-2}
\left. U^a \right|_{\mathrm{p}}=\frac{1}{\sqrt{1-\frac{u^2}{\left( 1-\frac{2M}{r} \right) ^2}}}\frac{1}{\sqrt{1-\frac{2M}{r}}}\left( 1,-u,0,0 \right),
\end{equation}
into the frequency Eq. (\ref{eq1-1}) yields the frequency of light at the point point $\mathrm{p}$,
\begin{equation}\label{eq2-3}
\omega _{\mathrm{p}} =\frac{\omega}{\sqrt{1-\frac{2M}{r'}}}\sqrt{\frac{1-\frac{2M}{r'}-u}{1-\frac{2M}{r'}+u}},
\end{equation}
In mind of the frequency transformation Eq. (\ref{eq2-0}), the gravitational redshift equation,
\begin{equation}\label{eq2-5}
\frac{\omega _{\mathrm{p}}}{\omega _{\mathrm{o}}} =\frac{1}{\sqrt{1-\frac{2M}{r}\sqrt{1-u^2}}}\sqrt{\frac{1-\frac{2M}{r}\sqrt{1-u^2}-u}{1-\frac{2M}{r}\sqrt{1-u^2}+u}},
\end{equation}
is derived by replacing $\omega$ with $\omega _{\mathrm{o}}$, and replacing $r'$ with $ r=r'\sqrt{1-u^2}$ the position of the black hole in coordinate system of a moving observer. Eq. (\ref{eq2-5}) describes the gravitational redshift of a light propagating in the positive direction $x$-axis with frequency $\omega _{\mathrm{o}}$, in the mean time, the Schwarzschild black hole moves toward the negative direction of $x$-axis at a velocity of $u$. A point $p$ that locates at a distance $r$ to the black hole measures a frequency of light with $\omega _{\mathrm{p}}$. Fig. \ref{redshift0} showed the redshift, $z = {\omega _{\mathrm{p}}}/{\omega _{\mathrm{o}}}$, of a light with respect to different distances $r$ and incoming velocity $u$. The light experienced redshift when when a black hole approaches to the observer and blueshift when it moves away from the observer.

\begin{figure}[htbp]
\centering
\par
\begin{center}
$
\begin{array}{c}
  \includegraphics[scale=0.48]{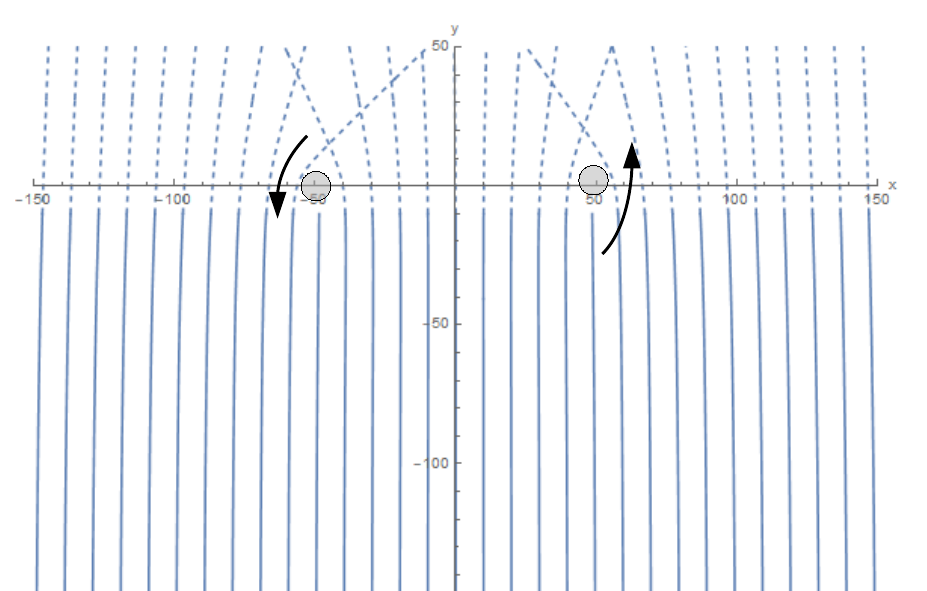}    \vspace{0.3in}\\
\mbox{\bf (a)}\\
\includegraphics[scale=0.4]{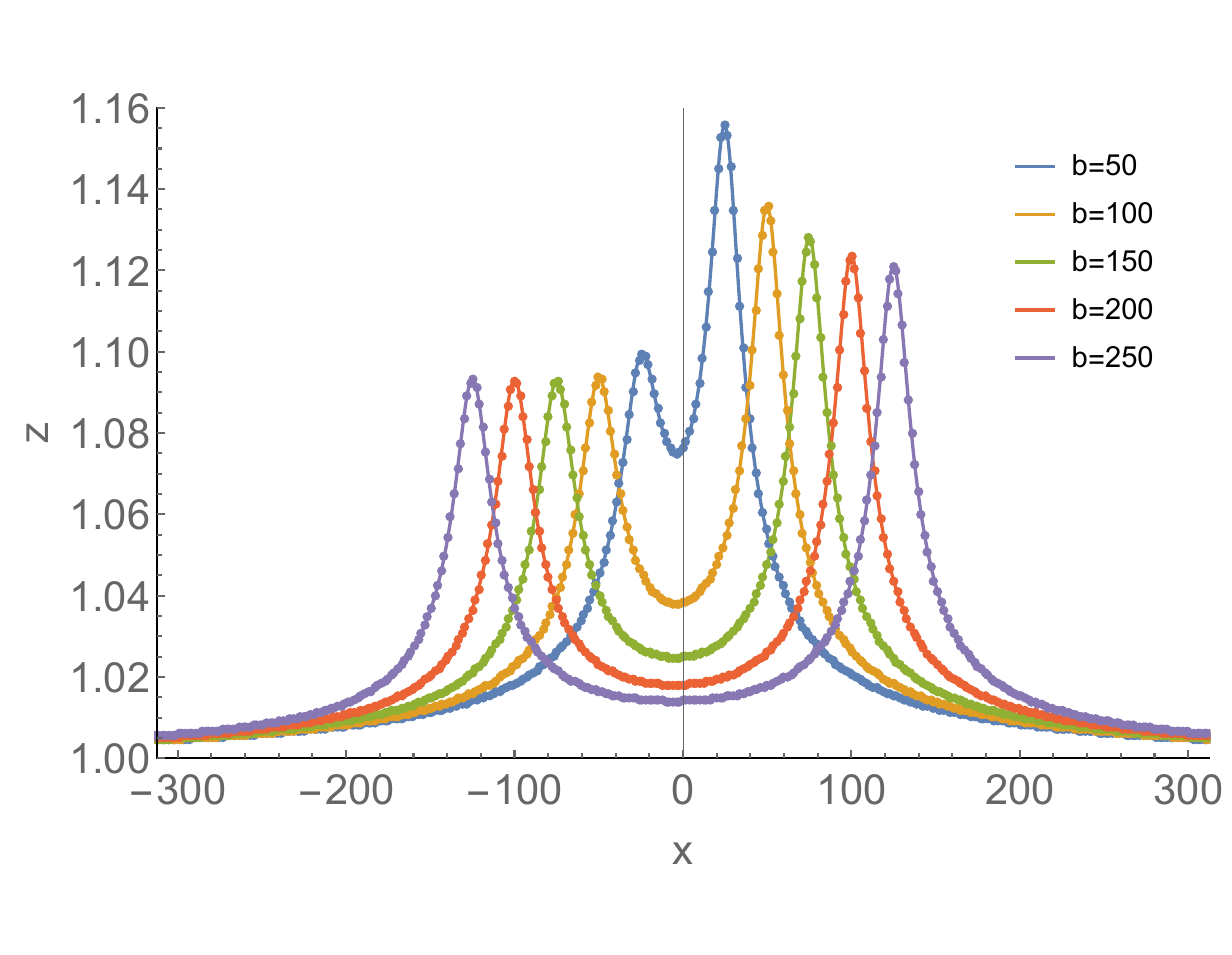}
  \\
\mbox{\bf (b)}
\end{array}
$
\end{center}
\caption{(a) The parallel geodesic lines from the negative infinity reaches $y = -10$. (b) The distribution of the redshift $z = \omega{'}/\omega$ on the $y = -10$ line with respect to $x$. $b$ is the distance between the two black holes}\label{redshift1}
\end{figure}

The gravitational redshift equation above holds for single black hole. The redshift factor on apparent horizons of binary black hole was numerical simulated for quasicircular binary inspirals \cite{3-1}. Here we calculated the gravitational redshift based on an approximated solution of the binary black hole \cite{Alvi}. For two black holes circling around their common center of mass, the curved spacetime around a binary black hole is described by an approximated solution \cite{Alvi}. The frequencies of a space time point in the centroid system is derived from the geodesic equation, $\omega =-K^{\mu}Z^{\nu}g_{\mu \nu}$, where $Z^a=\left( 1/\sqrt{-g_{00}},0,0,0 \right) $ is the velocity of the stationary observer in the frame of center-of-mass. The wave vector $K^{\mu}$ is the tangent vector of the geodesic line. The gravitation redshift is defined by the frequency ration of $\omega$ at a point in the far region to
to $\omega'$ in vicinity of the binary black hole, $ z=\omega'/\omega$. The incoming waves are travelling along parallel geodesic lines from the negative infinity of $y$-axis, propagating toward the positive direction of the $y$-axis and reaching the $x$-axis at $t = 0$. The two black holes locate exactly at the $x$-axis in Fig. \ref{redshift1} (a). The geodesic lines near the event horizon bends into different directions around the two rotating black holes in Fig. \ref{redshift1} (a), where the two black holes are located at $x = \pm 50$ and rotates in counterclockwise direction. The distribution of the gravitational redshift around the two rotating black holes at time $t=-10$ are showed in Fig. \ref{redshift1} (b). The black hole at $x = - 50$ moves towards the light source, inducing a smaller blue shift than that induced by the black hole at $x = 50$ which moves away from the light source.

Fig. \ref{redshift1} (b) showed the gravitational redshift which is numerically calculated by geodesic equation and the approximated solution of binary black hole \cite{Alvi}. The left and the right maximal redshift in Fig. \ref{redshift1} (b) are 1.094 and 1.136 which are generated by two black holes separated by a distance of $b = 50$. Two approximated gravitational redshift values are also derived, $z=1.098$ and $z=1.138$, by substituting the velocity of the left black hole $u=0.071$ (which moves towards the light source) and the velocity of the right black hole $u=-0.07$ (which moves away from the light source) into Eq. (\ref{eq2-5}). Therefore the gravitation redshift Eq. (\ref{eq2-5}) is consistent with the numerical result. The gravitational redshift on the left and the right side are not symmetrically distributed due to the rotation of the binary black holes. When the distance between two black holes grows from $b = 50$ to $b = 250$, the gravitational redshift gradually decays (Fig. \ref{redshift1} (b)).

\begin{figure}
  \centering
  \vspace{0.3in}
  \includegraphics[scale = 0.4]{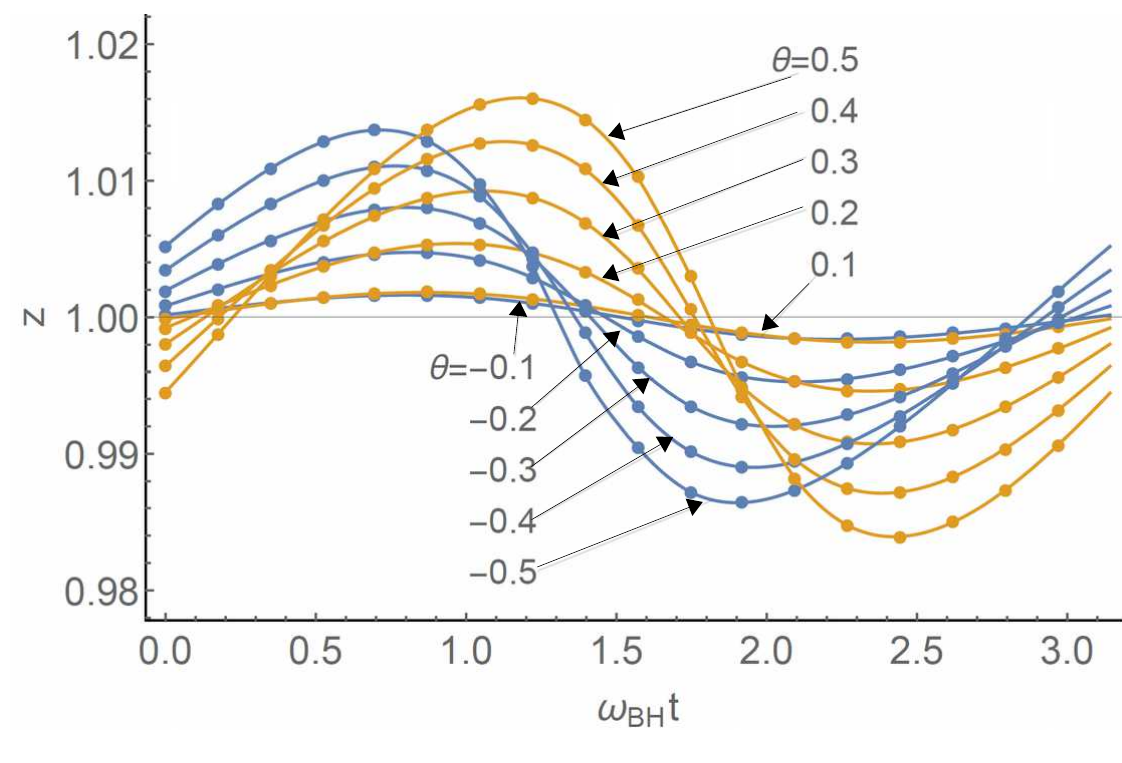}\\
  \caption{The evolution of redshift $z = \omega{'}/\omega$ around binary black hole within half a period. The blue line represents is light passing through the negative zone $x<0$. The red line represents lights passing through the positive zone $x>0$. Different curves represents light rays with different deflection angles.}\label{redshift3}
\end{figure}

The gravitational redshift oscillates periodically from redshift to blueshift (or vice versa) with respect to the rotating binary black hole (Fig.\ref{redshift3}). For fixed locations of the light source, the binary black hole and the Earth, each oscillating curve of redshift in Fig.\ref{redshift3} represents the redshift of a light deflected by an angle of $\theta$, i.e., the angle between the outgoing beam and the perpendicular line to the bonding vector of the binary black hole. The blue curves represent the light from the negative $x$ zone ($x<-50$), and the red curves represent the light from the positive $x$ zone $x>50$. The redshift is zero at zero deflection angle $\theta\rightarrow0$ in Fig. \ref{redshift3}, i.e, $z\rightarrow1$. The redshift grows with respect to a growing  deflection angle. The maximum redshift is of 1\% at $\theta=0.4$. The deflection angle of gravitational lens is usually very small in astronomical observations \cite{0-6,0-7}.

\section{Conclusions}

Propagating waves inference one another when curved geodesics of a binary black hole intersect with one another. The interference fringes of binary black holes has a bright central stripe accompanied by two dark adjacent stripes. Unlike the interference pattern of electromagnetic waves passing thorough double slits, the relative intensity of the bright and dark strips in the interference pattern of lights passing through binary black hole does not decease to zero, instead it preserves to infinity, providing a promising observation on Earth. As the binary black hole rotates, the bright central stripe turn into dark strip, while the adjacent dark stripes remain dark all the time. The average interference fringes is obtained by overlaying the interference fringes within an orbital period. Comparing with single black hole, the bright central stripe concentrates more energy, and the peak intensity of the central bright stripe is always larger than that of single black hole. When the two black holes rotate around their common center of mass, the two black holes result in opposite gravitation redshift, i.e., the gravitational redshift (blueshift) due to the left (right) black hole. The angular velocity of the binary black hole determines the asymmetric distribution of gravitational redshift. The amplitude of redshift is positively correlated to the deflection angle of incoming light. Therefore, the angular velocity of a binary black hole is measurable based on the observation of spatial distribution of gravitational redshift.

\end{document}